\documentclass[10pt,twoside]{article}
\usepackage{graphicx}
\usepackage{amsmath}
\usepackage{Latex-document}

\title{\bf Affine Weyl Group Approach \vskip -2mm
to Painlev\'e Equations \vskip 6mm}

\author{M. Noumi\vspace*{-0.5cm}\thanks{Department of Mathematics, School of Science and Technology, Kobe
University, Rokko, Kobe 657-8501, Japan. E-mail: noumi@math.kobe-u.ac.jp}}
\date{\vspace{-8mm}}

\usepackage{amssymb}
\newtheorem{theorem}{Theorem}

\newcommand{\comment}[1]{}
\newcommand{\PL}[1]{P_{\mbox{\scriptsize\sc #1}}}
\newcommand{\PN}[1]{N_{\mbox{\scriptsize\sc #1}}}
\newcommand{\HS}[1]{H_{\mbox{\scriptsize\sc #1}}}
\newcommand{\mat}[1]{
\arraycolsep=2pt \left[\begin{matrix} #1
\end{matrix}\right]
}
\newcommand{\diag}[1]{\mbox{\rm diag}({#1})}

\newcommand{\bb}{\mbox{\boldmath $b$}}

\newcommand{\bu}{\mbox{\boldmath $u$}}

\newcommand{\bep}{\mbox{\boldmath $\varepsilon$}}
\newcommand{\bphi}{\mbox{\boldmath $\varphi$}}
\newcommand{\brho}{\mbox{\boldmath $\rho$}}
\newcommand{\bkap}{\mbox{\boldmath $\kappa$}}

\begin{document}
\maketitle

\thispagestyle{first} \setcounter{page}{497}

\begin{abstract}\vskip 3mm

An overview is given on recent developments in the affine Weyl
group approach to Painlev\'e equations and discrete Painlev\'e
equations, based on the joint work with Y.~Yamada and K.~Kajiwara.

\vskip 4.5mm


\end{abstract}

\vskip 12mm

\section{Introduction}
\label{section 1}\setzero \vskip-5mm \hspace{5mm }

The purpose of this paper is to give a survey on recent
developments in the affine Weyl group approach to Painlev\'e
equations and discrete Painlev\'e equations.

It is known that each of the Painlev\'e equations from $\PL{II}$
through $\PL{VI}$ admits the action of an affine Weyl group as a
group of B\"acklund transformations (see a series of works
\cite{O} by K.~Okamoto, for instance). Furthermore, the B\"acklund
transformations (or the Schlesinger transformations) for the
Painlev\'e equations can already be thought of as discrete
Painlev\'e equations with respect to the parameters. The main idea
of the affine Weyl group approach to (discrete) Painlev\'e systems
is to extend this class of Weyl group actions to general root
systems, and to make use of them as the common underlying
structure that unifies various types of discrete
system\,(\cite{NYAff}). In this paper, we discuss several aspects
of affine Weyl group symmetry in nonlinear systems, based on a
series of joint works with Y.~Yamada and K.~Kajiwara.

Before starting the discussion of (discrete) Painlev\'e equations,
we recall some definitions, following the notation of \cite{K}. A
{\em $($generalized\,$)$ Cartan matrix\,} is an integer matrix
$A=(a_{ij})_{i,j\in I}$ (with a finite indexing set) satisfying
the conditions
\begin{equation}
a_{ii}=2; \quad a_{ij}\le 0\quad (i\ne j);\quad
a_{ij}=0\Longleftrightarrow a_{ji}=0.
\end{equation}
The {\em Weyl group} $W(A)$ associated with $A$ is defined by the
generators  $s_i$ ($i\in I$), called the {\em simple reflections},
and the fundamental relations
\begin{equation}
s_i^2=1,\qquad (s_is_j)^{m_{ij}}=1\quad(i\ne j),
\end{equation}
where $m_{ij}=2,3,4,6$ or $\infty$, according as
$a_{ij}a_{ji}=0,1,2,3$ or $\ge 4$. When the Cartan matrix
$A=(a_{ij})_{i,j=0}^l$ is of {\em affine type} (of type
$A^{(1)}_l$,$B^{(1)}_l$,\ldots,$D^{(3)}_4$), the corresponding
Weyl group is called an {\em affine Weyl group}. \comment{ A
characteristic feature of an affine Weyl group is that $W$ can be
expressed as the semidirect product $M\rtimes W_0$ of a free
abelian subgroup $M\simeq \mathbb{Z}^l$ and the finite Weyl group
$W_0=\langle s_1,\ldots,s_l\rangle$ acting on $M$. This fact is a
key in applying affine Weyl groups to the construction of discrete
systems. If an affine Weyl group $W$ is realized as a group of
birational transformations on some affine space $X$, then the
lattice part $M$ provides a commuting set of discrete flows on $X$
that is {\em covariant} under the action of the finite Weyl group
$W_0$ (see \cite{NYAff}). }

We fix some notation for the case of type $A^{(1)}_l$ that will be
used throughout this paper. The Cartan matrix
$A=(a_{ij})_{i,j=0}^{l}$ of type $A^{(1)}_l$ is defied by
\begin{equation}
A=\mat{2 & -2\cr -2 & 2}\ \ (l=1),\qquad A=\mat{ 2&-1& &&-1\cr
\vspace{-6pt} -1&2&-1&&\cr \vspace{-6pt} &-1&2&\ddots&\cr
&&\ddots&\ddots &-1\cr -1&&&-1 &2 } \ \ (l\ge 2).
\end{equation}
The affine Weyl group $W(A^{(1)}_{l})=\langle
s_0,s_1,\ldots,s_{l}\rangle$ is defined by the following
fundamental relations:
\begin{equation}
\begin{array}{rl}\smallskip
(l=1): & s_0^2=s_1^2=1,\cr (l\ge2):& s_i^2=1,\quad s_is_j=s_js_i\
\ (j\ne i,i\pm 1),\quad (s_is_j)^3=1\ \  (j=i\pm1),
\end{array}
\end{equation}
where we have identified the indexing set with
$\mathbb{Z}/(l+1)\mathbb{Z}$. We also define an extension
$\widetilde{W}(A^{(1)}_l)=\langle s_0,\ldots,s_{l}, \pi\rangle$ of
$W(A^{(1)}_l)$ by adjoining a generator $\pi$ (rotation of
indices) such that $\pi s_i=s_{i+1}\pi$  for all $i=0,1,\ldots,l$;
we do {\em not} impose the relation $\pi^{l+1}=1$. \comment{ This
group $\widetilde{W}=\widetilde{W}(A^{(1)}_l)$ decomposes into the
semidirect product $L\rtimes W_0$, where $L$ is the free abelian
group of rank $l+1$ generated by
\begin{equation}
\gamma_k=s_{k-1}\cdots s_1\pi s_l \cdots s_k
\quad(k=1,\ldots,l+1),
\end{equation}
and $W_0=\langle s_1,\ldots,s_l\rangle$ is isomorphic to the
symmetric group $S_{l+1}$ of degree $l+1$. }

\section{Variations on the theme of {\boldmath $P_{\mbox{\small IV}}$}}
\label{section 2}\setzero \vskip-5mm \hspace{5mm }

In this section, we present several examples of affine Weyl group
action of type $A^{(1)}_2$ to illustrate the role of affine Weyl
group symmetry in (discrete) Painlev\'e equations and related
integrable systems.

\subsection{Symmetric form of {\boldmath $\PL{\mbox{\bf IV}}$}}
 \vskip-5mm \hspace{5mm }

Consider the following system of nonlinear differential equations
for three unknown functions $\varphi_j=\varphi_j(t)$ ($j=0,1,2$):
\begin{equation}\label{eq:SymP4}
\renewcommand{\arraystretch}{1.2}
(\PN{IV})\qquad \left\{
\begin{array}{cc}
\varphi_0'=\varphi_0(\varphi_1-\varphi_2)+\alpha_0,\cr
\varphi_1'=\varphi_1(\varphi_2-\varphi_0)+\alpha_1,\cr
\varphi_2'=\varphi_2(\varphi_0-\varphi_1)+\alpha_2,
\end{array}
\right.
\end{equation}
where $'=d/dt$ denotes the derivative with respect to the
independent variable $t$, and $\alpha_j=0$ ($j=0,1,2$) are
parameters. When $\alpha_0+\alpha_1+\alpha_2=0$, this system
provides an integrable deformation of the Lotka-Volterra
competition model for three species. When
$\alpha_0+\alpha_1+\alpha_2=k\ne 0$, it is essentially the {\em
fourth Painlev\'e equation}
\begin{equation}
(\PL{IV})\qquad y''=\frac{1}{2y}(y')^2+\frac{3}{2}y^3+4ty^2+
(t^2-\alpha)y+\frac{\beta}{y}.
\end{equation}
In fact, from $(\varphi_0+\varphi_1+\varphi_2)'=k$, we have
$\varphi_0+\varphi_1+\varphi_2=kt+c$. Under the renormalization
$k=1$, $c=0$, system \eqref{eq:SymP4} can be written as a second
order equation for $y=\varphi_0$; it is transformed into $\PL{IV}$
with $\alpha=\alpha_2-\alpha_1$, $\beta=-2\alpha_0^2$ by the
change of variables $t\to\sqrt{2}t$, $y\to -y/\sqrt{2}$. In view
of this fact, we call \eqref{eq:SymP4} the {\em symmetric form} of
the fourth Painlev\'e equation ($\PN{IV}$). This type of
representation for $\PL{IV}$ was introduced by \cite{VS}, \cite{A}
in the context of nonlinear dressing chains, and by \cite{NYP4} in
the study of rational solutions of $\PL{IV}$.

The symmetric form $\PN{IV}$ provides a convenient framework for
describing the discrete symmetry of $\PL{IV}$. Let
$\mathcal{K}=\mathbb{C}(\alpha,\varphi)$ be the field of rational
functions in the variables $\alpha=(\alpha_0,\alpha_1,\alpha_2)$
and $\varphi=(\varphi_0,\varphi_1, \varphi_2)$. We define the
derivation $\ ':\mathcal{K}\to\mathcal{K}$ by using formulas
\eqref{eq:SymP4} together with $\alpha_j'=0$ ($j=0,1,2$); we
regard the differential field $(\mathcal{K}, \ ')$ as representing
the differential system $\PN{IV}$. In this setting, we say that an
automorphism of $\mathcal{K}$ is a {\em B\"acklund
transformation\/} for $\PN{IV}$ if it commutes with the derivation
$'$. (A B\"acklund transformation as defined above means a
birational transformation of the phase space that commutes with
the flow defined by the nonlinear differential system.) As we will
see below, $\PN{IV}$ has four fundamental B\"acklund
transformations that generate the extended affine Weyl group
$\widetilde{W}=\langle s_0,s_1,s_2,\pi\rangle$ of type
$A^{(1)}_2$. Identifying the indexing set $\{0,1,2\}$ with
$\mathbb{Z}/3\mathbb{Z}$, we define the automorphisms $s_i$
($i=0,1,2$) and $\pi$ of $\mathcal{K}$ by
\begin{equation}\label{eq:addW}
\begin{array}{lll}\smallskip
s_i(\alpha_j)=\alpha_j-\alpha_i a_{ij}, \quad&
s_i(\varphi_j)=\varphi_j+\dfrac{\alpha_i}{\varphi_i} u_{ij}
\quad&(i,j=0,1,2),\cr
\pi(\alpha_j)=\alpha_{j+1},\quad&\pi(\varphi_j)=\varphi_{j+1}
&(j=0,1,2).
\end{array}
\end{equation}
Here $A=(a_{ij})_{i,j=0}^2$ stands for the Cartan matrix of type
$A^{(1)}_2$, and $U=(u_{ij})_{i,j=0}^2$ for the orientation matrix
of the Dynkin diagram (triangle) in the positive direction:
\begin{equation}
A=\mat{2&-1&-1\cr-1&2&-1\cr-1&-1&2}, \qquad
U=\mat{0&1&-1\cr-1&0&1\cr1&-1&0}.
\end{equation}
These automorphisms $s_i$ and $\pi$ commute with the derivation
$'$, and satisfy the fundamental relations
\begin{equation}
s_i^2=1, \quad (s_is_{i+1})^3=1 ,\quad \pi s_i=s_{i+1}\pi\qquad
(i=0,1,2)
\end{equation}
for the generators of $\widetilde{W}(A^{(1)}_2)$. Hence we obtain
a realization of the extended affine Weyl group
$\widetilde{W}(A^{(1)}_2)$ as a group of B\"acklund
transformations for $\PN{IV}$. Notice that the action of the
affine Weyl group $W=\langle s_0,s_1,s_2\rangle$ on the
$\alpha$-variables is identical to its canonical action on the
{\em simple roots}.

We remark that the affine Weyl group symmetry is deeply related to
the structure of special solutions of $\PL{IV}$ (with the
parameters $\alpha_j$ as in $\PN{IV}$). Along each reflection
hyperplane $\alpha_j=n$ ($j=0,1,2$; $n\in \mathbb{Z}$) in the
parameter space, $\PL{IV}$ has a one-parameter family of classical
solutions expressed in terms of Toeplitz determinants of
Hermite-Weber functions; each solution of this class is obtained
by B\"acklund transformations from a seed solution at $\alpha_j=0$
which satisfies a Riccati equation. Also, at each point of the
$W$-orbit of the barycenter
$(\alpha_0,\alpha_1,\alpha_2)=(\frac{1}{3},\frac{1}{3},\frac{1}{3})$
of the fundamental alcove, it has a rational solution expressed in
terms of Jacobi-Trudi determinants of Hermite polynomials.

\subsection{{\boldmath $q$}-Difference analogue of {\boldmath $\PL{\bf IV}$}}
 \vskip-5mm \hspace{5mm }

We now introduce a {\em multiplicative} analogue of the birational
realization \eqref{eq:addW} of the extended affine Weyl group
$\widetilde{W}=\langle s_0,s_1,s_2,\pi\rangle$ (\cite{KNYqP4}).
Taking the field of rational functions
$\mathcal{L}=\mathbb{C}(a,f)$ in the variables $a=(a_0,a_1,a_2)$
and $f=(f_0,f_1,f_2)$, we define the automorphisms $s_0, s_1, s_2,
\pi$ of $\mathcal{L}$ as follows:
\begin{equation}\label{eq:mulW}
\renewcommand{\arraystretch}{1.4}
\begin{array}{lll}
s_i(a_j)=a_j a_i^{-a_{ij}},
&s_i(f_j)=f_j\left(\dfrac{a_i+f_i}{1+a_if_j}\right)^{u_{ij}}
&(i,j=0,1,2),\cr \pi(a_j)=a_{j+1},\quad &\pi(f_j)=f_{j+1}
&(j=0,1,2),
\end{array}
\end{equation}
where $a_j$ are the multiplicative parameters corresponding to the
simple roots $\alpha_j$. These automorphisms again satisfy the
fundamental relations for the generators of $\widetilde{W}$. In
the following, the $\widetilde{W}$-invariant $a_0a_1a_2=q$ plays
the role of the base for $q$-difference equations. If one
parameterizes $a_j$ and $f_j$ as
\begin{equation}\label{eq:ftophi}
a_j=e^{-\varepsilon^2\alpha_j/2},\quad f_j=-e^{-\varepsilon
\varphi_j} \qquad(j=0,1,2)
\end{equation}
with a small parameter $\varepsilon$, one can recover the original
formulas \eqref{eq:addW} from \eqref{eq:mulW} by taking the limit
$\varepsilon\to 0$.

A $q$-difference analogue of (the symmetric form of) $\PL{IV}$ is
given by
\begin{equation}\label{eq:qP4}
\renewcommand{\arraystretch}{1.4}
(q\PL{IV})\qquad \left\{
\begin{array}{ll}
\smallskip
T(f_0)= a_0a_1f_1 \dfrac{1+a_2f_2+a_2a_0f_2f_0}
{1+a_0f_0+a_0a_1f_0f_1}, \cr
\smallskip
T(f_1)= a_1a_2f_2 \dfrac{1+a_0f_0+a_0a_1f_0f_1}
{1+a_1f_1+a_1a_2f_1f_2}, \cr T(f_2)= a_2a_0f_0
\dfrac{1+a_1f_1+a_1a_2f_1f_2} {1+a_2f_2+a_2a_0f_2f_0}, \cr
T(a_j)=a_j\qquad(j=0,1,2),
\end{array}
\right.
\end{equation}
where $T$ stands for the discrete time evolution (\cite{KNYqP4}).
Notice that \eqref{eq:qP4} implies $T(f_0f_1f_2)=(a_0a_1a_2)^2
f_0f_1f_2 =q^2 f_0f_1f_2$; hence one can consistently introduce a
time variable $t$ such that $f_0f_1f_2=t^2$. If we consider $f_j$
as functions of $t$, the discrete time evolution $T$ is identified
with the $q$-shift operator $t\to q\,t$, so that
$Tf_j(t)=f_j(q\,t)$. In this sense, formula \eqref{eq:qP4} defines
a system of nonlinear $q$-difference equations, which we call the
{\em fourth $q$-Painlev\'e equation} ($q\PL{IV}$).

The time evolution $T$, regarded as an automorphism of
$\mathcal{L}$, commutes with the action of $\widetilde{W}$ that we
already described above. Namely, the $q$-difference system
$q\PL{IV}$ admits the action of the extended affine Weyl group
$\widetilde{W}$ as a group of B\"acklund transformations. Again,
by taking the limit as $\varepsilon\to 0$ under the
parametrization \eqref{eq:ftophi}, one can show that the
$q$-difference system $q\PL{IV}$, as well as its affine Weyl group
symmetry, reproduces the differential system $\PN{IV}$. It is
known that $q\PL{IV}$ defined above shares many characteristic
properties with the original $\PL{IV}$. For example, it has
classical solutions expressed by continuous $q$-Hermite-Weber
functions, and rational solutions expressed by of continuous
$q$-Hermite polynomials, analogously to the case of $\PL{IV}$
(\cite{KNYqP4}, \cite{KNYqKP}). We also remark that, when
$a_0a_1a_2=1$, one can regard $q\PL{IV}$ as a discrete integrable
system which generalizes a discrete version of the Lotka-Volterra
equation.

\subsection{Ultra-discretization of {\boldmath $\PL{\bf IV}$}}
 \vskip-5mm \hspace{5mm }

It should be noticed that the discrete time evolution of
$q\PL{IV}$ is defined in terms of a {\em subtraction-free}
birational transformation; we say that a rational function is
subtraction-free if it can be expressed as a ratio of two
polynomials with real positive coefficients. Recall that there is
a standard procedure, called the {\em ultra-discretization}, of
passing from subtraction-free rational functions to piecewise
linear functions (\cite{TTMS}, \cite{BFZ}, see also
\cite{NYtRSK}). Roughly, it is the procedure of replacing the
operations
\begin{equation}
a\cdot b \to A+B,\quad a/b \to A-B, \quad a+b\to \max(A,B).
\end{equation}
\comment{ Under the parametrization $a=e^{A/\varepsilon}$,
$b=e^{B/\varepsilon}$, the passage from $a+b$ to $\max(A,B)$ is
understood formally through the limiting procedure
$\underset{\varepsilon\to+0}{\lim} {\varepsilon}
\log(e^{A/\varepsilon}+e^{B/\varepsilon})=\max(A,B)$. }

Introducing the variables $A_j$, $F_j$ ($j=0,1,2$), from
$q\PL{IV}$ we obtain the following system of piecewise linear
difference equations by ultra-discretization:
\begin{equation}
\arraycolsep=2pt (u\PL{IV})\ \ \left\{\begin{array}{lll}
T(F_0)=&A_0+A_1+F_1+\max(0,A_2+F_2,A_2+A_0+F_2+F_0)\cr
\smallskip
&\quad-\max(0,A_0+F_0,A_0+A_1+F_0+F_1), \cr
T(F_1)=&A_1+A_2+F_2+\max(0,A_0+F_0,A_0+A_1+F_0+F_1)\cr
\smallskip
&\quad-\max(0,A_1+F_1,A_1+A_2+F_1+F_2), \cr
T(F_2)=&A_2+A_0+F_0+\max(0,A_1+F_1,A_1+A_2+F_1+F_2)\cr
\smallskip
&\quad-\max(0,A_2+F_2,A_2+A_0+F_2+F_0), \cr
T(A_j)=&A_j\qquad(j=0,1,2),
\end{array}
\right.
\end{equation}
which we call the {\em fourth ultra-discrete Painlev\'e equation}
($u\PL{IV}$). Simultaneously, the affine Weyl group symmetry of
$q\PL{IV}$ is ultra-discretized as follows:
\begin{equation}
\arraycolsep=4pt
\begin{array}{ll}\smallskip
s_i(A_j)=A_j-A_i a_{ij},& s_i(F_j)=F_j+u_{ij}\big(
\max(A_i,F_i)-\max(0,A_i+F_i) \big),\cr \pi(A_j)=A_{j+1},&
\pi(F_j)=F_{j+1}\qquad(i,j=0,1,2).
\end{array}
\end{equation}
This time, the extended affine Weyl group $\widetilde{W}$ is
realized as a group of piecewise linear transformations on the
affine space with coordinates $(A,F)$. We also remark that, when
$A_0+A_1+A_2=Q=0$, $u\PL{IV}$ gives rise to an ultra-discrete
integrable system. It would be an interesting problem to analyze
special solutions of the ultra-discrete system $u\PL{IV}$.

\section{Discrete symmetry of Painlev\'e equations}
\label{section 3} \setzero\vskip-5mm \hspace{5mm}

In this section, we propose a uniform description of discrete
symmetry of the Painlev\'e equations $P_{J}$ for $J=\mbox{\small
II, IV, V, VI}$.  We also give some remarks on a generalization of
this class of birational Weyl group action to arbitrary root
systems.

\subsection{Hamiltonian system {\boldmath $H_J$}}
 \vskip-5mm \hspace{5mm }

It is known that each Painlev\'e equation $P_J$ ($J=\mbox{\small
II, III,\ldots, VI}$) is equivalently expressed as a Hamiltonian
system
\begin{equation}
(H_J):\qquad \frac{dq}{dt}= \frac{\partial H}{\partial p},\quad
\frac{dp}{dt}=-\frac{\partial H}{\partial q}
\end{equation}
with a polynomial Hamiltonian $H=H(q,p,t,\alpha)\in
\mathbb{C}(t)[q,p,\alpha]$ depending on parameters
$\alpha=(\alpha_1,\ldots,\alpha_l)$ (see \cite{IKSY}, for
instance). Setting $\mathcal{K}=\mathbb{C}(q,p,t,\alpha)$, we
define the Poisson bracket
$\{\cdot,\cdot\}:\mathcal{K}\times\mathcal{K}\to\mathcal{K}$ and
the Hamiltonian vector field $\delta:\mathcal{K}\to\mathcal{K}$ by
\begin{equation}
\{\varphi,\psi\}= \dfrac{\partial \varphi}{\partial
p}\dfrac{\partial \psi}{\partial q} -\dfrac{\partial
\varphi}{\partial q}\dfrac{\partial \psi}{\partial p}, \quad
\delta(\varphi)=\{H,\varphi\}+\frac{\partial\varphi}{\partial t}
\qquad(\varphi,\psi\in\mathcal{K}).
\end{equation}
In this setting, a B\"acklund transformation for $H_J$ is
understood as an automorphism $w:\mathcal{K}\to\mathcal{K}$
that 
commutes with $\delta$. We also say that $w$ is {\em canonical\,}
if it preserves the Poisson bracket:
$w(\{\varphi,\psi\})=\{w(\varphi),w(\psi)\}$ for any
$\varphi,\psi\in\mathcal{K}$.

For each $J=\mbox{\small II, III,\ldots,VI}$, it is known that the
parameter space for $H_J$ is identified with the Cartan subalgebra
of a semisimple Lie algebra, and that an extension of the
corresponding affine Weyl group acts on $\mathcal{K}$ as a group
of B\"acklund transformations (\cite{O}). A table of fundamental
B\"acklund transformations for $H_J$ can be found in \cite{NTY}.

If the Hamiltonian $H$ is chosen appropriately, the affine Weyl
group symmetry of $H_J$ for $J=\mbox{\small II, IV, V, VI}$ can be
described in a universal way in terms of root systems. With the
notation of \cite{K}, the type of the affine root system is
specified as follows\footnote{ In the case of $\HS{III}$, one can
use an extension of the affine Weyl group, either of type
$C^{(1)}_2$ or of $2A^{(1)}_1$, for describing the same group of
B\"acklund transformations. It seems natural to expect that the
same principle to be discussed below should apply to $\HS{III}$ as
well, but we have not completely understood the case of $\HS{III}$
yet.}.
\begin{equation}
\renewcommand{\arraystretch}{1.4}
\begin{array}{c|cccc}
H_J & \HS{II}  & \HS{IV} & \HS{V} & \HS{VI} \cr \hline
X^{(1)}_l&A^{(1)}_1 & A^{(1)}_2 & A^{(1)}_3 & D^{(1)}_4
\end{array}
\end{equation}
The corresponding Cartan matrix $A=(a_{ij})_{i,j=0}^l$ is given by
\begin{equation}
\begin{array}{rcrc}
\medskip
A^{(1)}_1: & A=\mat{2 & -2 \cr -2 & 2}\quad& A^{(1)}_2: & A=\mat{2
& -1 & -1 \cr -1 & 2 & -1\cr -1&-1&2} \cr A^{(1)}_3: & A=\mat{2 &
-1 &0  &-1 \cr
  -1 & 2 & -1 & 0 \cr
0 & -1 & 2 & -1  \cr -1 & 0 & -1 & 2}\quad& D^{(1)}_4: & A=\mat{2
& 0 & -1 & 0 & 0 \cr 0 & 2 & -1 & 0 & 0\cr -1&-1&2&-1&-1\cr 0 &
0&-1&2&0\cr 0&0&-1&0&2}
\end{array}
\end{equation}
respectively. For the description of affine Weyl group symmetry,
we make use of the following Hamiltonian $H=H(q,p,t,\alpha)$:
\begin{equation}\label{eq:Ham}
\renewcommand{\arraystretch}{1.2}
\arraycolsep=2pt
\begin{array}{rrl}
\HS{II}: & H=&\dfrac{1}{2}p(p-2q^2+t)+\alpha_1 q, \cr \HS{IV}: &
H=&qp(2p-q-2t)-2\alpha_1p-\alpha_2q, \cr \HS{V}: &
tH=&q(q-1)p(p+t)-(\alpha_1+\alpha_3)qp +\alpha_1p+\alpha_2 t q,
\cr \HS{VI}: & t(t-1)H=&q(q-1)(q-t)p^2-\big\{(\alpha_0-1)q(q-1)
+\alpha_4(q-1)(q-t) \cr &&\quad +\alpha_3q(q-t)\big\}
p+\alpha_1(\alpha_1+\alpha_2)(q-t).
\end{array}
\end{equation}
The parameter $\alpha_0$ is defined so that
$\alpha_0+\alpha_1+\cdots+\alpha_l=1$ for $J=\mbox{\small II, IV,
V}$, and $\alpha_0+\alpha_1+2\alpha_2+\alpha_3+\alpha_4=1$ for
$J=\mbox{\small VI}$. (Conventionally, the null root is normalized
to be the constant $1$.)

\subsection{Discrete symmetry of {\boldmath $H_J$}}
 \vskip-5mm \hspace{5mm }

Our main observation concerning the discrete symmetry of $H_J$ can
be summarized as follows.
\begin{theorem} \label{thm1}
Choose the polynomial Hamiltonian $H\in\mathbb{C}(t)[q,p,\alpha]$
for $H_{J}$ $(J=\mbox{\small\rm II, IV, V, VI})$ as in
\eqref{eq:Ham}. Then there exists a set
$\{\varphi_0,\varphi_1,\ldots,\varphi_l\}$ of nonzero elements in
the Poisson algebra $\mathcal{R}=\mathbb{C}[q,p,t]$ with the
following properties\,$:$
\newline
$(1)$ The elements $\varphi_i$ $(i=0,1,\ldots,l)$ satisfy the
Serre relations of type $X^{(1)}_l$
\begin{equation}
\mbox{\em ad}_{\{\}}(\varphi_i)^{-a_{ij}+1}\varphi_j=0 \qquad(i\ne
j),
\end{equation}
where $\mbox{\rm ad}_{\{\}}(\varphi)=\{\varphi,\cdot\}$ stands for
the adjoint action of $\varphi$ by the Poisson bracket.
\newline
$(2)$  For each $i=0,1,\ldots,l$, define $s_i$ to be the unique
automorphism of $\mathcal{K}=\mathbb{C}(q,p,t,\alpha)$ such that
\begin{equation}
\begin{array}{ll}
\smallskip
s_i(\alpha_j)=\alpha_j-\alpha_i a_{ij} & (j=0,1,\ldots,l),\cr
s_i(\psi)=\exp\big( \dfrac{\alpha_i}{\varphi_i}\mbox{\rm
ad}_{\{\}}(\varphi_i)\big) \psi \quad& (\psi\in
\mathcal{R}=\mathbb{C}[q,p,t]).
\end{array}
\end{equation}
Then these $s_i$ are canonical B\"acklund transformations for
$H_J$. Furthermore, the subgroup $W=\langle
s_0,s_1,\ldots,s_l\rangle$ of $\mbox{\rm Aut}(\mathcal{K})$ is
isomorphic to the affine Weyl group $W(X^{(1)}_l)$.
\end{theorem}

Note that, for each $\psi\in\mathcal{R}$, $s_i(\psi)$ is
determined as a finite sum
\begin{equation}
s_i(\psi)=\psi+ \dfrac{\alpha_i}{\varphi_i}\{\varphi_i,\psi\}+
\dfrac{1}{2!}\big(\dfrac{\alpha_i}{\varphi_i}\big)^2
\{\varphi_i,\{\varphi_i,\psi\}\}+\cdots,
\end{equation}
since the action of $\mbox{ad}_{\{\}}(\varphi_i)$ on $\mathcal{R}$
is locally nilpotent. A choice of the generators $\varphi_i$
($i=0,1,2,\ldots,l$) with the properties of Theorem \ref{thm1} is
given as follows:
\begin{equation}
\renewcommand{\arraystretch}{1.2}
\begin{array}{rl}
\HS{II}: & \varphi_0=-p+2q^2+t,\quad\varphi_1=p. \cr \HS{IV}: &
\varphi_0=-p+\dfrac{q}{2}+t,\quad \varphi_1=-\dfrac{q}{2},\quad
\varphi_2=p. \cr \HS{V}: &
\varphi_0=p+t,\quad\varphi_1=tq,\quad\varphi_2=-p,
\quad\varphi_3=t(1-q). \cr \HS{VI}: & \varphi_0=q-t,\quad
\varphi_1=1,\quad \varphi_2=-p,\quad \varphi_3=q-1,\quad
\varphi_4=q.
\end{array}
\end{equation}
We remark that, in the case of $H_J$ ($J=\mbox{\small II, IV, V}$)
of type $A^{(1)}_l$ ($l=1,2,3$), we also have the B\"acklund
transformation $\pi$ corresponding to the diagram rotation; its
action is given simply by $\pi(\alpha_j)=\alpha_{j+1}$,
$\pi(\varphi_j)=\varphi_{j+1}$. \comment{ For those who wish to
inspect examples, we give below the table of fundamental
B\"acklund transformations for $\HS{II}$.
\begin{equation}
\renewcommand{\arraystretch}{1.4}
\begin{array}{c||cc|ccc}
& \alpha_0 & \alpha_1 & q & p & t\cr \hline\hline s_0 & -\alpha_0&
\alpha_1+2\alpha_0 & q +\frac{\alpha_0}{p-2q^2-t} &
p+\frac{4\alpha_0 q}{p-2q^2-t}+\frac{2\alpha_0^2}{(p-2q^2-t)^2} &
t \cr s_1 & \alpha_0+2\alpha_1 & -\alpha_1 & q+\frac{\alpha_1}{p}
& p & t \cr \hline \pi & \alpha_1 & \alpha_0 & -q & -p+2q^2+t & t
\end{array}
\end{equation}
}

If we use the polynomials $\varphi_j$ as dependent variables, the
Hamiltonian system $\HS{IV}$, for example, is rewritten as
\begin{equation}
\frac{d\varphi_j}{dt}=2 \varphi_j(\varphi_{j+1}-\varphi_{j+2})+
\alpha_j\qquad(j=0,1,2)
\end{equation}
with the convention $\varphi_{j+3}=\varphi_j$, from which we
obtain the symmetric form $\PN{IV}$ by a simple rescaling of the
variables. We remark that the polynomials $\varphi_j$ are the
factors of the ``leading term'' of the Hamiltonian $H$. Also, in
the context of irreducibility of Painlev\'e equations, the
polynomials $\varphi_j$ are the fundamental {\em invariant
divisors} along the reflection hyperplanes $\alpha_j=0$ (see
\cite{NO}, \cite{UW}, for instance). When $\alpha_j=0$, the
specialization of $H_J$ by $\varphi_j=0$  gives rise to a Riccati
equation that reduces to a linear equation of hypergeometric type;
for $J=\mbox{\small II, IV, V and VI}$, the differential equations
of Airy, Hermite-Weber, Kummer and Gauss appear in this way,
respectively.

Apart from differential equations, this class of birational
realization of Weyl groups as in Theorem \ref{thm1} can be
formulated for an arbitrary Cartan matrix by means of Poisson
algebras (see \cite{NYBir}, for the details). In this sense,
B\"acklund transformations for Painlev\'e equations $P_J$
($J=\mbox{\small II, IV, V, VI}$) have a {\em universal\/} nature
with respect to root systems. In the case where $A$ is of affine
type, such a birational realization of the affine Weyl group
appears as the symmetry of systems of nonlinear partial
differential equations of Painlev\'e type, obtained by similarity
reduction from the principal Drinfeld-Sokolov hierarchy\,(of
modified type). The case of type $A^{(1)}_l$ will be mentioned in
the next section. As for the original Painlev\'e equations,
$\PL{II}$, $\PL{IV}$ and $\PL{V}$ are in fact obtained by
similarity reduction from the $(l+1)$-reduced modified KP
hierarchy for $l=1,2,3$, respectively. For $\PL{VI}$, an $8\times
8$ Lax pair is constructed in \cite{NYLaxP6} in the framework of
the affine Lie algebra $\widehat{\mathfrak{so}}(8)$. This Lax pair
is compatible with the affine Weyl group symmetry of Theorem
\ref{thm1}. It is not clear, however, how this construction should
be understood in relation to the Drinfeld- Sokolov hierarchy of
type $D^{(1)}_4$.

\section{Painlev\'e systems with {\boldmath $W(A^{(1)}_l)$} symmetry}
\label{section 4} \setzero\vskip-5mm \hspace{5mm }

In this section, we introduce Painlev\'e systems and
$q$-Painlev\'e systems with affine Weyl group symmetry of type
$A$; this part can be regarded as a generalization, to higher rank
cases, of the variations of $\PL{IV}$ discussed in Section 2. In
the following, we fix two positive integers $M, N$, and consider a
Painlev\'e system, as well as its $q$-version, attached to
$(M,N)$.

\subsection{Painlev\'e system of type {\boldmath $(M,N)$}}
 \vskip-5mm \hspace{5mm }

We investigate the compatibility condition for a system of linear
differential equations
\begin{equation}\label{eq:Lin}
N z\partial_z\vec{\psi}=A\vec{\psi},\quad
\partial_{t_m}\vec{\psi}=B_m \vec{\psi}\quad(m=1,\ldots,M),
\end{equation}
where $\vec{\psi}=(\psi_1,\ldots,\psi_N)^{\mbox{\rm\small t}}$ is
the column vector of unknown functions, and $A$, $B_m$ are
$N\times N$ matrices, both depending on
$(z,t)=(z,t_1,\ldots,t_M)$. We assume that
\begin{equation}
B_m=\sum_{k=0}^{m-1} \diag{\bb^{(m,k)}}\Lambda^k+\Lambda^m,
\quad(m=1,\ldots,M),
\end{equation}
where $\Lambda=\sum_{i=1}^{N-1} E_{i,i+1}+z E_{N,1}$ denotes the
cyclic matrix, $E_{ij}=(\delta_{a,i}\delta_{b,j})_{a,b}$ being the
matrix units, and $\bb^{(m,k)}=(b^{(m,k)}_1,\ldots,b^{(m,k)}_N)$
are $N$-vectors depending only on $t$. Note that the compatibility
condition
\begin{equation}\label{eq:ZS}
\partial_{t_n}(B_m)-\partial_{t_m}(B_n)+
[B_m,B_n]=0 \qquad(m,n=1,\ldots,M)
\end{equation}
is the Zakharov-Shabat equation of the $N$-reduced modified KP
hierarchy (restricted to the first $M$ time variables). As for the
matrix $A$, we set
\begin{equation}
A=-\diag{\brho}+\sum_{k=1}^M kt_k B_k, \quad
\brho=(N-1,N-2,\ldots,0).
\end{equation}
Then the compatibility condition
\begin{equation}\label{eq:AB}
\partial_{t_m}(A)-N z\partial_{z}(B_m)+[A,B_m]=0
\qquad(m=1,\ldots,M),
\end{equation}
reduces to the homogeneity condition
\begin{equation}\label{eq:Sim}
\sum_{n=1}^M nt_n\partial_{t_n}(\bb^{(m,k)}) =(k-m) \bb^{(m,k)}
\quad(1\le k\le m\le M)
\end{equation}
for the coefficients of the $B$ matrices. We define the {\em
Painlev\'e system of type\,} $(M,N)$ to be the system of nonlinear
partial differential equations \eqref{eq:ZS} with the similarity
constraint \eqref{eq:Sim}.

We remark that, when $(M,N)=(3,2), (2,3), (2,4)$, this system
reduces essentially to the Painlev\'e equations $\PL{II}, \PL{IV},
\PL{V}$, respectively. When $(M,N)=(2,N)$ ($N\ge 2$), it
corresponds to the higher order Painlev\'e equation of type
$A^{(1)}_{N-1}$ discussed in \cite{NYHigh}. Note that the linear
problem \eqref{eq:Lin} defines a monodromy preserving deformation
of linear ordinary differential system of order $N$ on
$\mathbb{P}^1$, with one regular singularity at $z=0$ and one
irregular singularity at $z=\infty$.

The Painlev\'e system of type $(M,N)$ admits the action of the
affine Weyl group $W=\langle s_0,s_1,\ldots,s_{N-1}\rangle$ of
type $A^{(1)}_{N-1}$ as a group of B\"acklund transformations.
Expressing the matrix $A$ as
\begin{equation}
-A=\diag{\brho}-\sum_{k=1}^{M} kB_k =\diag{\bep}+\sum_{k=1}^{M}
\diag{\bphi^{(k)}} \Lambda^k,
\end{equation}
we set $\alpha_0=N-\varepsilon_1+\varepsilon_N$,
$\alpha_i=\varepsilon_i-\varepsilon_{i+1}$ and
$\varphi_0=\varphi^{(1)}_N$, $\varphi_i=\varphi^{(1)}_i$ for
$i=1,\ldots,N-1$. (Note that all the exponents $-\varepsilon_i$ at
$z=0$ are constant.) Then, for each $i=0,\ldots,N-1$, the
B\"acklund transformation $s_i$ is obtained as the compatibility
of the gauge transformation
\begin{equation}
s_i\vec{\psi}=G_i\vec{\psi},\quad
G_i=1+\dfrac{\alpha_i}{\varphi_i} F_i \qquad(i=0,1,\ldots,N-1),
\end{equation}
where $F_0=z^{-1}E_{1,N}$ and $F_i=E_{i+1,i}$ ($i=1,\ldots,N-1$).
If we regard $\alpha_i$ and $\varphi^{(k)}_i$ as coordinates for
the matrix $-A$, the ring $\mathcal{R}$ of polynomials in the
$\varphi$-variables has a natural structure of Poisson algebra. In
these coordinates, the B\"acklund transformation $s_i$ is
determined by the universal formula
\begin{equation}
s_i(\alpha_j)=\alpha_j-\alpha_i a_{ij}, \quad s_i(\psi)=\exp\big(
\dfrac{\alpha_i}{\varphi_i}\mbox{\rm ad}_{\{\}}(\varphi_i)\big)
\psi  \quad (\psi\in \mathcal{R}).
\end{equation}

\subsection{{\boldmath $q$}-Painlev\'e system of type {\boldmath $(M,N)$}}
 \vskip-5mm \hspace{5mm }

We construct the $q$-Painlev\'e system of type $(M,N)$ in an
analogous way (\cite{KNYqKP}). With the time variables $t_m$ and
the $q$-shift operators $T_m=T_{q,t_m}$ ($i=1,\ldots,M$), we
investigate the compatibility condition for a system of linear
$q$-difference equations
\begin{equation} \label{eq:qLin}
T_{q^N,z}\vec{\psi}=A \vec{\psi}, \quad
T_{m}\vec{\psi}=B_m\vec{\psi} \quad (m=1,\ldots,M),
\end{equation}
where $A$, $B$ are $N\times N$ matrices depending on $(z,t)$. We
assume that
\begin{equation}
B_m=\diag{\bu^{(m)}}+t_m\Lambda, \quad
\bu^{(m)}=(u^{(m)}_1,\ldots,u^{(m)}_N) \quad(m=1,\ldots,M),
\end{equation}
with compatibility condition
\begin{equation}\label{eq:qZS}
T_n(B_m)B_n=T_m(B_n)B_m\qquad(m,n=1,\ldots,M).
\end{equation}
We consider this condition as the Zakharov-Shabat equation for the
$N$-reduced modified $q$-KP hierarchy; in this formulation, all
the time variables $t_1,\ldots,t_M$ are treated equally. Note
that, as to the Euler operator $T=T_1\cdots T_M$, we have
\begin{equation}
T\vec{\psi}=B_T\vec{\psi}, \quad B_T=T_2\cdots T_M(B_1) T_3\cdots
T_M(B_2)\cdots T_{M}(B_{M-1}) B_M.
\end{equation}
In the linear $q$-difference system \eqref{eq:qLin}, we choose the
following matrix for $A$:
\begin{equation}
A=\diag{\bkap}^{-1}\ B_T,\quad \bkap=(q^{N-1},q^{N-2},\ldots,1).
\end{equation}
Then the compatibility condition
\begin{equation}
T_m(A) B_m = T_{q^N,z}(B_m) A\qquad(m=1,\ldots,M)
\end{equation}
is equivalent to the homogeneity condition
\begin{equation}\label{eq:qSim}
T_1\cdots T_M(u^{(m)}_i)=u^{(m)}_i \qquad(i=1,\ldots,N;
m=1,\ldots,M).
\end{equation}
We define the {\em $q$-Painlev\'e system of type} $(M,N)$ to be
the system of nonlinear $q$-difference equations \eqref{eq:qZS}
for $M\times N$ unknown functions $u^{(m)}_i$ ($m=1,\ldots,M$;
$i=1,\ldots,N$) with the similarity constraint \eqref{eq:qSim}.
This system can be written in the form
\begin{equation}
T_m(u^{(n)}_i)=F^{(m,n)}_i(t,u) \qquad(m,n=1,\ldots,M; i=1,\ldots
N);
\end{equation}
in general, these $F^{(m,n)}_i(t,u)$ are complicated rational
functions. It turns out, however, that by introducing new
variables
\begin{equation}\label{eq:xx}
x^i_{j}=\frac{1}{t_i} T_{i+1}T_{i+2}\cdots T_{M}(u^{(i)}_j)
\quad(i=1,\ldots,M; j=1,\ldots,N),
\end{equation}
the time evolution of the $q$-Painlev\'e system can be described
explicitly by means of a birational affine Weyl group action on
the $x$-variables.

\subsection{A birational Weyl group action
on the matrix space}
 \vskip-5mm \hspace{5mm }

For describing the time evolution $T_i$ of the $q$-Painlev\'e
system, we introduce a birational action of the direct product
$\widetilde{W}(A^{(1)}_{M-1}) \times \widetilde{W}(A^{(1)}_{N-1})
$ of two extended affine Weyl groups. In the following, we use the
notation
\begin{equation}
\widetilde{W}^{M} =\langle r_0,r_1,\ldots,r_{M-1},\omega\rangle,
\quad \widetilde{W}_N=\langle s_0,s_1,\ldots,s_{N-1},\pi\rangle
\end{equation}
for the two extend affine Weyl groups. Introducing two parameters
$q$, $p$, we take $\mathbb{K}=\mathbb{C}(q,p)$ as the ground
field. Let $\mathcal{K}=\mathbb{K}(x)$ be the field of rational
functions in the $M N$ variables $x^i_j$ ($1\le i\le M; 1\le j\le
N$); we regard the $x$-variables as the canonical coordinates of
the affine space of $M\times N$ matrices. For convenience, we
extend the indices $i$,$j$ of $x^i_j$ to $\mathbb{Z}$ by setting
$x^{i+M}_j=qx^i_j$, $x^i_{j+N}=px^i_j$.

We define the automorphisms $r_k$ ($k\in\mathbb{Z}/M\mathbb{Z}$),
$\omega$, $s_l$   ($l\in\mathbb{Z}/N\mathbb{Z}$), $\pi$ of
$\mathcal{K}$ as follows:
\begin{equation}
\begin{array}{ll}
\smallskip
r_k(x^i_j)=px^{i+1}_j\dfrac{P^i_{j-1}}{P^i_j}, \quad
r_k(x^{i+1}_j)=p^{-1}x^{i}_j\dfrac{P^i_{j}}{P^i_{j-1}}
\quad(i\equiv k \mod M); \cr\smallskip r_k(x^i_j)=x^i_j \qquad
(i\not\equiv k \mod M); \quad \omega(x^i_j)=x^{i+1}_j; \cr
\smallskip
s_l(x^i_j)=qx^{i}_{j+1}\dfrac{Q^{i-1}_{j}}{Q^i_j}, \quad
s_l(x^{i}_{j+1})=q^{-1}x^{i}_j\dfrac{Q^{i}_{j}}{Q^{i-1}_{j}}
\quad(j\equiv l \mod N); \cr s_l(x^i_{j})=x^i_j \qquad
(j\not\equiv l \mod N); \quad \pi(x^i_j)=x^{i}_{j+1},
\end{array}
\end{equation}
where
\begin{equation}
P^i_j={\displaystyle\sum_{k=1}^N \prod_{a=0}^{k-1}\ x^i_{j+a}
\prod_{a=k+1}^{N}\ x^{i+1}_{j+a}}, \quad
Q^i_j={\displaystyle\sum_{k=1}^M \prod_{a=0}^{k-1}\ x^{i+a}_j
\prod_{a=k+1}^{M}\ x^{i+a}_{j+1}}.
\end{equation}
\comment{
\begin{equation}
\begin{array}{ll}
\smallskip
P^i_j={\displaystyle\sum_{k=1}^N}\ x^i_j\cdots x^i_{j+k-1}
x^{i+1}_{j+k+1}\cdots x^{i+1}_{j+N}, \cr
Q^i_j={\displaystyle\sum_{k=1}^M}\ x^i_j\cdots x^{i+k-1}_{j}
x^{i+k+1}_{j+1}\cdots x^{i+M}_{j+1}.
\end{array}
\end{equation}
} Note that all these automorphisms represent subtraction-free
birational transformations on the affine space of $M\times N$
matrices.

\begin{theorem}
The automorphisms $r_0,\ldots,r_{M-1},\omega$ and
$s_0,\ldots,s_{M-1},\pi$ of $\mathcal{K}$ defined as above give a
realization of the product $\widetilde{W}^M\times\widetilde{W}_N$
of extended affine Weyl groups.
\end{theorem}

By using this birational action of affine Weyl group
$\widetilde{W}^M=\langle r_0,\ldots,r_{M-1},\omega\rangle$ we
define $\gamma_1,\ldots, \gamma_M$ by
\begin{equation}\label{eq:gam}
\gamma_k=r_{k-1}\cdots r_{1}\,\omega\, r_{M}\cdots r_{k}
\qquad(k=1,\ldots,M).
\end{equation}
We remark that these elements $\gamma_k$ generate a free abelian
subgroup $L\simeq \mathbb{Z}^M$, and that the extended affine Weyl
group $\widetilde{W}^M$ decomposes into the semidirect product
$L\rtimes S_M$ of $L$ and the symmetric group of degree $M$ that
acts on $L$ by permuting the indices for $\gamma_k$.

\begin{theorem}
In terms of the variables $x^i_j$ defined by \eqref{eq:xx}, the
time evolution of the $q$-Painlev\'e system of type $(M,N)$ is
described by
\begin{equation}
T_k(x^i_j)=\gamma^{-1}_k(x^i_j)\quad(1\le i\le M;1\le j\le N)
\end{equation}
for all $k=1,\ldots,M$, where $\gamma_k$ is defined by
\eqref{eq:gam} through the birational action of $\widetilde{W}^M$
with $p=1$.
\end{theorem}

This theorem means that the discrete time evolutions $T_k$
($k=1,\ldots,M$) of the $q$-Painlev\'e system of type $(M,N)$
coincides with the commuting discrete flows $\gamma_k^{-1}$
($k=1,\ldots,M$) arising from the affine Weyl group action of
$\widetilde{W}^M$ with $p=1$. Furthermore the $q$-Painlev\'e
system admits the action of extended affine Weyl group
$\widetilde{W}_N$ of type $A^{(1)}_{N-1}$ as a group of B\"acklund
transformations. One can show that the fourth $q$-Painlev\'e
equation $q\PL{IV}$ discussed in Section 2 arises from the
$q$-Painlev\'e system of type $(M,N)=(2,3)$, consistently with the
differential case.

Finally we give some remarks on the ultra-discretization. From the birational action of $\widetilde{W}^{M}\times
\widetilde{W}_N$ with two parameters $q,p$, we obtain a piecewise linear action of the same group on the space of
$M\times N$ matrices, with two parameters $Q,P$ corresponding to $q,p$. When $P=0$, the commuting piecewise linear
flows $\gamma_k\in\widetilde{W}^M$ may be called the {\em ultra-discrete Painlev\'e system of type} $(M,N)$. When
$P=Q=0$, it specializes to an integrable ultra-discrete system; it gives rise to an $M$-periodic version of the
box-ball system.

This class of piecewise linear action is tightly related to the
combinatorics of {\em crystal bases}. The coordinates of the
$M\times N$ matrix space can be identified with the coordinates
for the tensor product $\mathcal{B}^{\otimes M}$ of $M$ copies of
the crystal basis $\mathcal{B}$ for the symmetric tensor
representation of $GL_N$. Under this identification, it turns out
that the piecewise linear transformations $r_k$ and $s_l$ with
$P=Q=0$ represent the action of the combinatorial $R$ matrices and
the Kashiwara's Weyl group action on $\mathcal{B}^{\otimes M}$,
respectively (see \cite{NYtRSK}).

\label{lastpage}

\end{document}